\documentclass[multphys,vecarrow]{svmult}
\usepackage{subfigure}

\usepackage{graphicx}
\usepackage{amssymb}
\usepackage{amsmath}
\usepackage{float}
\usepackage{subeqnar}
\usepackage{epsfig}

\begin{document}
\frontmatter
\mainmatter

\hyphenation{Glei-chungs-sys-tem set-ups}
 
\bibliographystyle{unsrt}

\title*{Enhanced empirical data for the fundamental diagram and the
  flow through bottlenecks}
\titlerunning{Enhanced empirical data}
\author{A. Seyfried\inst{1}, M. Boltes\inst{1}, J. K\"ahler\inst{2}, 
W. Klingsch\inst{2}, A. Portz\inst{1},\\ 
T. Rupprecht\inst{2}, A. Schadschneider\inst{3}, B. Steffen\inst{1}, 
and A. Winkens\inst{2}}
\institute{
J\"ulich Supercomputing Centre -- Forschungszentrum J\"ulich GmbH\\
52425 J\"ulich -- Germany\\
e-mail: {\tt a.seyfried@fz-juelich.de}\\
\and
Institute for Building Material Technology and Fire Safety Science -- 
Bergische Universit\"at Wuppertal\\
Pauluskirchstrasse 11 -- 42285 Wuppertal -- Germany\\
e-mail: {\tt klingsch@uni-wuppertal.de}\\
\and
Institut f\"ur Theoretische Physik -- Universit\"at zu K\"oln\\
50937 K\"oln -- Germany\\
e-mail: {\tt as@thp.uni-koeln.de}\\
}
\authorrunning{Seyfried et al.}

\date{\today}

\maketitle

\begin{abstract}
In recent years, several approaches for modelling pedestrian dynamics
have been proposed and applied e.g. for design of egress routes. 
However, so far not much attention has been paid to their \emph{quantitative} 
validation. This unsatisfactory situation belongs amongst others on the 
uncertain and contradictory experimental data base. The fundamental diagram, 
i.e.\ the density-dependence of the flow or velocity, is probably the most 
important relation as it connects the basic parameter to describe the dynamic 
of crowds. But specifications in different handbooks as well as experimental 
measurements differ considerably. The same is true for the bottleneck
flow. After a comprehensive review of the experimental data base we
give an survey of a research project, including experiments with up to 250
persons performed under well controlled laboratory conditions. The
trajectories of each person are measured in high precision to analyze the
fundamental diagram and the flow through bottlenecks. The trajectories allow
to study how the way of measurement influences the resulting relations.
Surprisingly we found large deviation amongst the methods. These may be
responsible for the deviation in the literature mentioned above. The results 
are of particular importance for the comparison of experimental data gained 
in different contexts and for the validation of models. 
\\
\noindent
\end{abstract}

\section{Introduction}

The number of models for pedestrian dynamics has grown in the past
years, but the experimental data to test them and to discriminate
between these models is still to a large extent uncertain and
contradictory (see e.g.\ \cite{Schadschneider}). In most models, 
pedestrians are considered to be autonomous mobile agents, hopping 
particles in a cellular automaton or self-driven particles in a 
continuous space. If the objective is to make quantitative 
predictions, like evacuation or travel times, the model has to be 
calibrated with empirical data.

One of the most important characteristics of pedestrian dynamics is the 
fundamental diagram giving the relation between pedestrian flow and density. 
Beside its importance for the dimensioning of pedestrian facilities it is
associated with every qualitative self-organization phenomenon, like
the formation of lanes or the occurrence of congestions. However,
specifications of different experimental studies, guidelines and 
handbooks, all display non negligible differences even for the most
relevant characteristics like maximal flow 
values, the corresponding density and the density where the flow is 
expected to become zero due to overcrowding. The connection between 
fundamental diagram and bottleneck flow is important as well and not 
really understood. In particular the maxima of fundamental diagrams 
are significantly lower than maximal flow values measured at 
bottlenecks. 

Although a large variety of models for pedestrian dynamics
has been proposed, so far there have been only limited attempts
to calibrate and validate these approaches. One reason is the unclear 
situation of the empirical data, as described above. 
This situation is very unsatisfactory and poses serious limitations 
on the use of such models e.g.\ in the area of safety planning. 
To improve the current state of affairs it is necessary to have
more reliable data that can be used as basis for validation and
calibration which then would allow to make quantitative predictions 
based on computer simulations.


In Sec.~\ref{sec-emp} we give a review of empirical results and discuss 
their discrepancies by comparing various experimental data and specifications 
from the literatur. To resolve some of the contradictions we initiated  
a research project including experiments with up to 250 persons under well 
controlled laboratory conditions, see Sec.~\ref{res-pro}. Great emphasis was 
given to the method of data recording by video technique and careful preparation 
of the experimental setups. This enables the accurate determination of all 
trajectories providing a microscopic insight into pedestrian dynamics, see \cite{Boltes2008}. 
In Sec.~\ref{meas-meth} we analyse  how the measurement method influences the 
resulting outcomes. 

\section{Review of Empirical Results}
\label{sec-emp}

\subsection{Fundamental Diagram}

The fundamental diagram describes the empirical relation between
density $\rho$ and flow $J$ (or specific flow per unit width
$J_s=J/w$). The name already indicates its importance and naturally it
has been the subject of many investigations.  Due to the hydrodynamic
relation $J=\rho\,v\,w$ there are three equivalent forms: $J_s(\rho)$,
$v(\rho)$ and $v(J_s)$.  In applications the relation is a basic input
for engineering methods developed for the design and dimensioning of
pedestrian facilities \cite{Predtechenskii1978,Fruin1971,Weidmann1993,Nelson2002}.
In this section 
we will concentrate on planar facilities like sidewalks, corridors or halls.
For various facilities like floors, stairs or ramps
the shape of the diagrams differ, but in general it is assumed that the
fundamental diagrams for the same type of facilities but different widths
merge into one diagram for the specific flow $J_s$. 

\begin{figure}[thb]
  \centerline{
    \includegraphics[width=0.45\columnwidth]{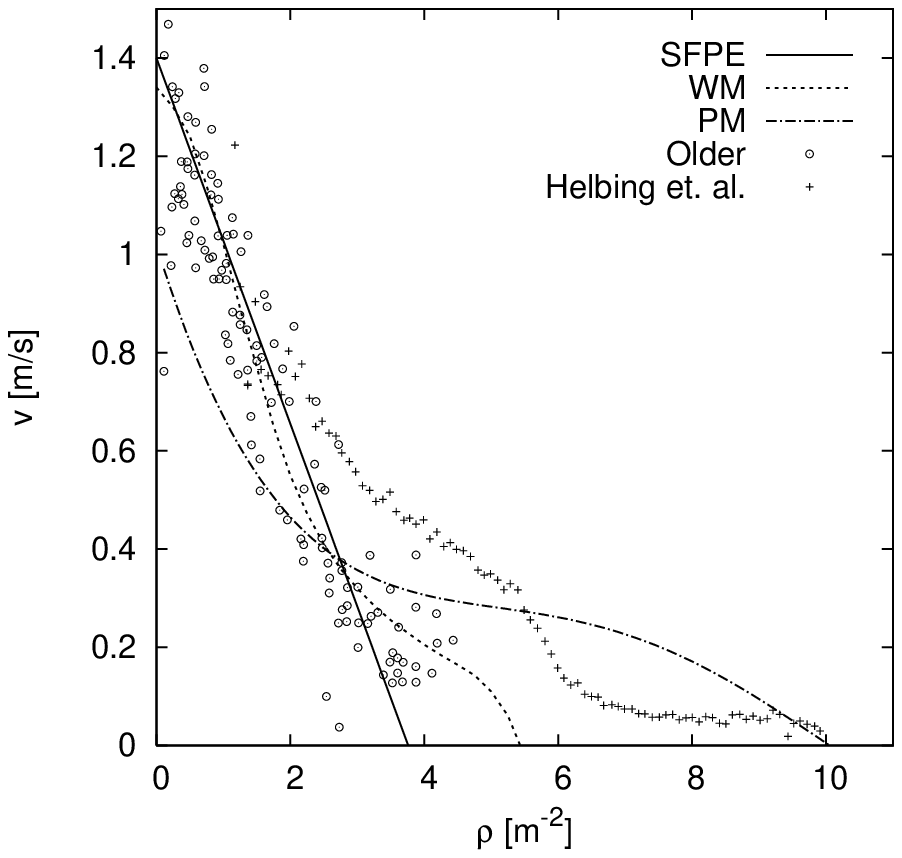}
    \hspace{0.3cm}
    \includegraphics[width=0.45\columnwidth]{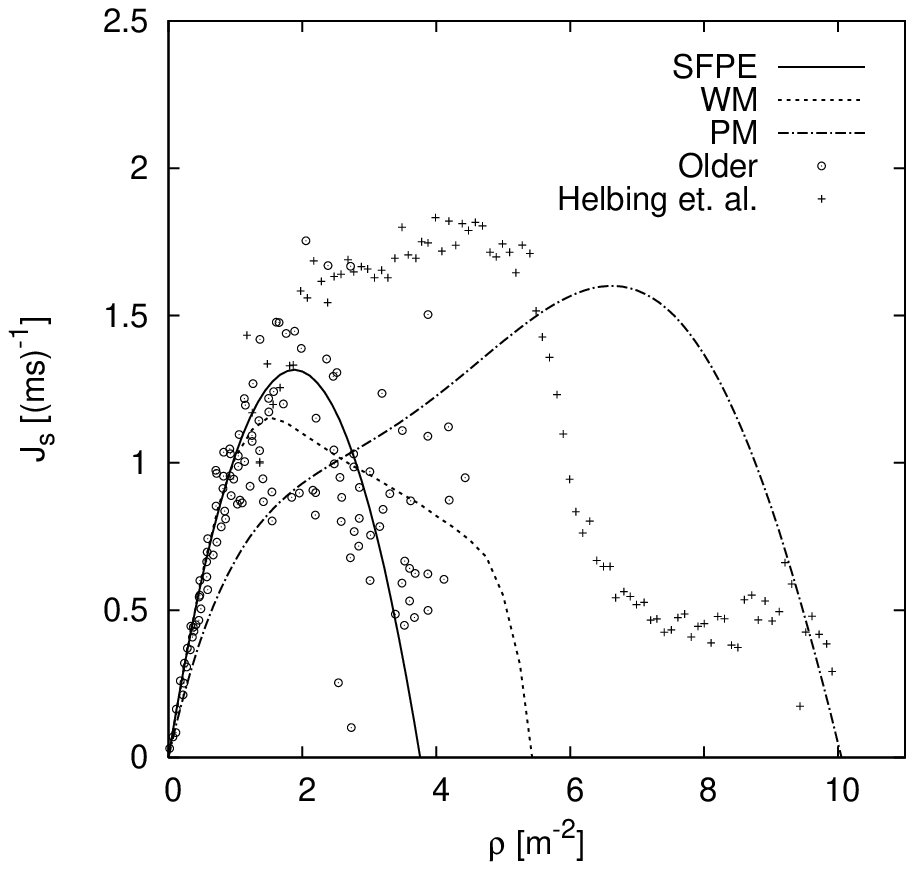}
  }
  \caption{Fundamental diagrams for pedestrian movement in planar facilities. 
    The lines refer to specifications according to planing guidelines
    (SFPE Handbook \cite{Nelson2002}, PM: Predtechenskii and Milinskii 
    \cite{Predtechenskii1978}, WM: Weidmann \cite{Weidmann1993}). Data points 
    give the range of experimental measurements \cite{Older1968,Helbing2007}.}
  \label{RHO-JS-EMP}
\end{figure}

Fig.~\ref{RHO-JS-EMP} shows various fundamental diagrams used in planing
guidelines plus the measurements of two selected empirical studies 
representing the overall range of the data. The comparison reveals
that specifications and measurements disagree considerably. In
particular the maximum of the function giving the capacity
$J_{s,{\rm max}}$ ranges from $1.2\;$(ms)$^{-1}$ to
$1.8\;$(ms)$^{-1}$, the density $\rho_0$ where the velocity approaches
zero due to overcrowding ranges from $3.8\;$m$^{-2}$ to
$10\;$m$^{-2}$ and, most notably, the density value where the maximum flow is
reached $\rho_c$ ranges from $1.75\;$m$^{-2}$ to $7\;$m$^{-2}$.
Several explanations for these deviations have been suggested,
including cultural and population differences
\cite{Helbing2007}, differences between uni- and
multidirectional flow \cite{Navin1969,Pushkarev1975},
short-ranged fluctuations \cite{Pushkarev1975}, influence of
psychological factors given by the incentive of the movement
\cite{Predtechenskii1978} and, partially related to the latter, the type of
traffic (commuters, shoppers) \cite{Oeding1963}.

The most elaborate fundamental diagram has been given by
Weidmann, who collected 25 data sets. An examination of the data which
were included in Weidmann's analysis shows that most measurements with
densities larger than $\rho=1.8\;$m$^{-2}$ are performed on
multidirectional streams.
Weidmann neglected differences between uni- and multidirectional flow
in accordance with Fruin, who states in his often cited book
\cite{Fruin1971} that the fundamental diagrams of multidirectional 
and unidirectional flow differ only slightly.  
This disagrees with results of Navin and Wheeler \cite{Navin1969} who
found a reduction of the flow in dependence of directional imbalances.
Here lane formation in bidirectional flow has to be considered. 
Bidirectional pedestrian flow includes
unordered streams as well as lane-separated and thus quasi-unidirectional
streams in opposite directions. Another explanation is given by Helbing 
et al.\ \cite{Helbing2007}
who argue that cultural and population differences are responsible
for the deviations between Weidmann and their data. In contrast to
this interpretation the data of Hanking and Wright \cite{Hankin1958}
gained by measurements in the London subway (UK) are in good agreement
with the data of Mori and Tsukaguchi \cite{Mori1987} measured in
the central business district of Osaka (Japan), both on strictly
uni-directional streams. This brief discussion clearly shows that up to
now there is no consensus about the origin of the discrepancies
between different fundamental diagrams and how one can explain the
shape of the function.

However, all diagrams agree in one characteristic: velocity decreases
with increasing density. As the discussion above indicates
there are many possible reasons and causes for the velocity reduction.
For the movement of pedestrians along a line a linear relation between
speed and the inverse of the density was measured in \cite{Seyfried2005}.
The speed for walking pedestrians depends also linearly on the step
size \cite{Weidmann1993} and the inverse of the density can be regarded as
the required length of one pedestrian to move. Thus it seems that smaller
step sizes caused by a reduction of the available space with
increasing density is, at least for a certain density region, one cause
for the decrease of speed. However, this is only a starting point
for a more elaborated modeling of the fundamental diagram.

%

\subsection{Bottleneck Flow} 
\label{BCK}

One of the most important practical questions is how the capacity of
the bottleneck increases with rising width. Studies of this dependence
can be traced back to the beginning of the last century
\cite{Dieckmann1911,Fischer1933} and are up to now discussed
controversially. 
At first sight, a stepwise increase of capacity with the width appears
to be natural if lanes are formed. For independent lanes, where 
pedestrians in one lane are not influenced by those in others, 
capacity increases only if an additional lane can be formed. 

In contrast, the study \cite{Seyfried2007} found that the distance of lanes 
and the speed in a lane increases with the bottleneck width until a new 
lane is formed, when the lanes come closer together again. 
This variation of lane distance leads to a very weak dependence of the 
density and velocity inside the bottleneck on its width. Thus in reference to 
$J=\rho\;v\;w$ the flow does not directly depend on the number of lanes. To 
find a conclusive judgement whether the capacity grows continuously with the 
width the results of different laboratory experiments 
\cite{Muller1981,Muir1996,Nagai2006,Kretz2006a,Seyfried2007c}
are compared in \cite{Seyfried2007c}.

\begin{figure}[thb]
  \centerline{
    \includegraphics[width=0.45\columnwidth]{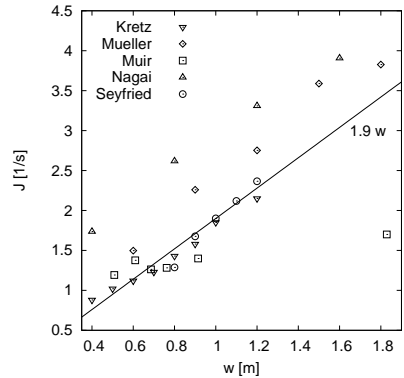}
  }
  \caption{Influence of the width of a bottleneck on the flow.
    Experimental data [M\"uller \cite{Muller1981}; Muir et al.~\cite{Muir1996}; 
    Nagai et al.~\cite{Nagai2006}; Seyfried et al.~\cite{Seyfried2007c}]
    of different types of bottlenecks and initial
    conditions. All data are taken under laboratory conditions where
    the test persons are advised to move normally.}
  \label{J-B-EMP}
\end{figure}

In the following we discuss the data of flow measurement collected in
Fig.~\ref{J-B-EMP}. The data by Muir et al.\ \cite{Muir1996}, who
studied the evacuation of airplanes, seem to support the stepwise
increase of the flow with the width. They show constant flow values
for $w > 0.6$~m.  But the flow there does not increase much up to
$w=1.8$~m, which indicates that in this special setup the flow is
limited by some other process, e.g. reaching the corridor. Thus all
data collected from flow measurements in Fig.~\ref{J-B-EMP} are
compatible with a continuous and almost linear increase with the
bottleneck width for $w>0.6\;m$.  Surprisingly the data in
Fig.~\ref{J-B-EMP} differ considerably in the values of the bottleneck
capacity. In particular the flow values of Nagai \cite{Nagai2006} and
M\"uller \cite{Muller1981} are much higher than the maxima of
empirical fundamental diagrams.  The comparison of the different
experimental setups shows that the exact geometry of the bottleneck is
of only minor influence on the flow while a high initial density in
front of the bottleneck can increase the resulting flow values.  This
leads to another interesting question, that is to say how the
bottleneck flow is connected to the fundamental diagram. General
results for driven diffusive systems \cite{Popkov1999} show that
boundary conditions only \emph{select} between the states of the
undisturbed system instead of creating completely different ones.
Therefore it is surprising that the measured maximal flow at
bottlenecks can exceed the maximum of the empirical fundamental
diagram. These questions are related to the common jamming criterion.
Generally it is assumed that a jam occurs if the incoming flow exceeds
the capacity of the bottleneck. In this case one expects the flow
through the bottleneck to continue with the capacity (or lower
values). The data presented in \cite{Seyfried2007c} show a more
complicated picture, we refer to the contribution of A. Winkens and T.
Rupprecht in these proceedings. While the density in front of the
bottleneck amounts to $\rho \approx 5.0~(\pm 2) \;$m$^{-2}$, the
density inside the bottleneck tunes around $\rho \approx
1.8\,$m$^{-2}$.

\section{Research Project - Overview}
\label{res-pro}

The research project is funded by the DFG and based on cooperation between the
Bergische Universit\"at Wuppertal, the Universit\"at zu K\"oln and
Forschungszentrum J\"ulich GmbH. It covers the execution of large scale
experiments, the data collection via automated determination of trajectories
with high accuracy, microscopic and macroscopic data analysis and the 
development of models to describe the dynamic of pedestrians quantitatively. 
In this section we give an overview of the experiments performed. 
  
As outlined in the previous section, there are a lot of possible
influences on the characteristics of pedestrian crowd movement. To
reduce as much as possible uncontrollable influences we decided to use
a homogenous group of test persons and to perform the experiments
under well controlled laboratory conditions. It is obvious that
results performed under special conditions are not suited for design
recommendations of e.g. escape routes. However such types of
experiments make it possible to study the influence of single
parameters, like the bottleneck width, and thus to resolve whether the
capacity of a bottleneck increases linearly or step wise.  Moreover
the determination of the trajectories of all persons with high
accuracy allows a microscopic insight into pedestrian dynamics and
thus to provide a secure data base for the development and microscopic
verification of models. Concerning the determination of trajectories
we refer to \cite{Boltes2008}.

\begin{figure}[thb]
  \centerline{
    \includegraphics[width=0.37\columnwidth]{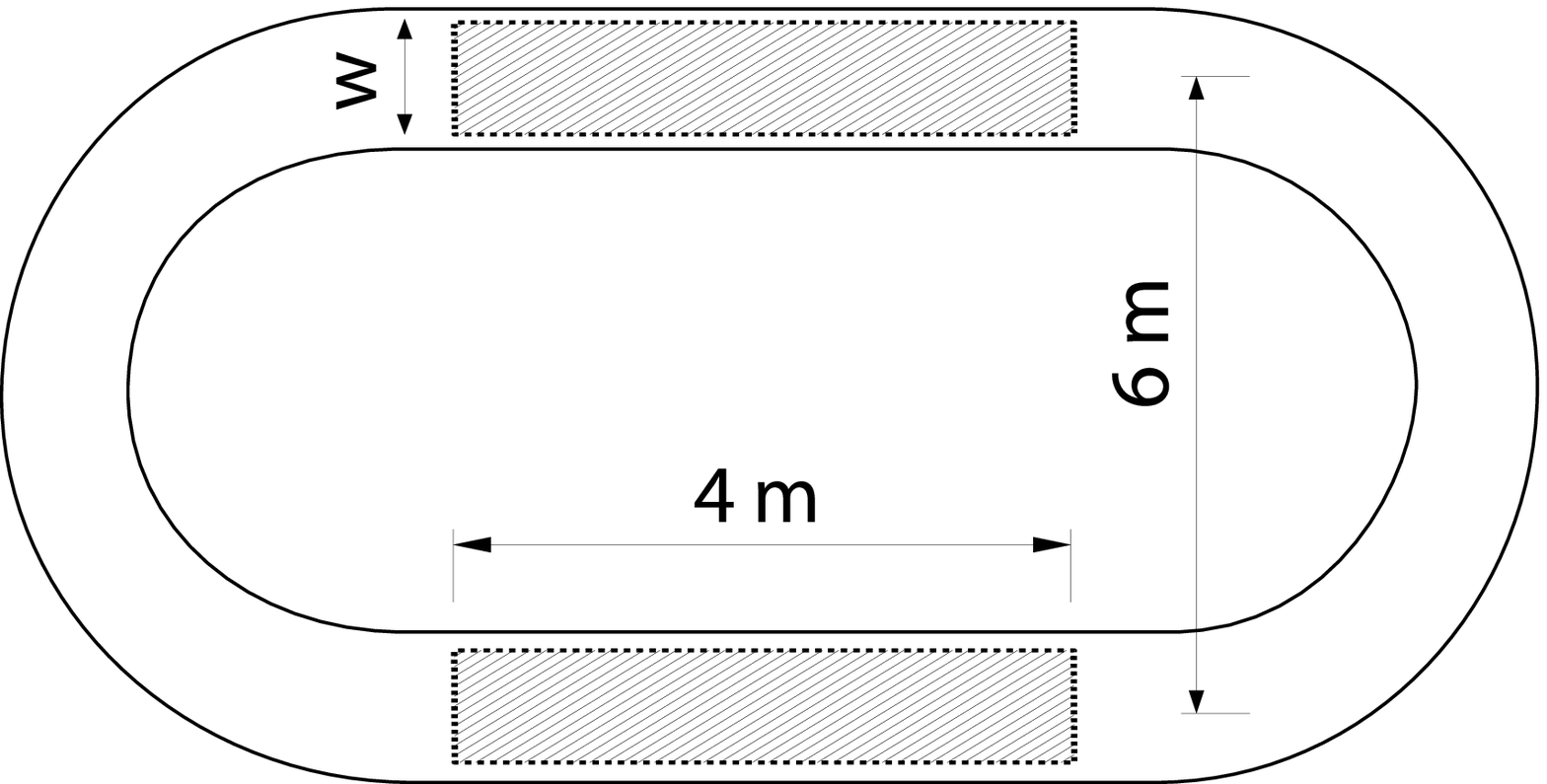}
    \qquad
    \includegraphics[width=0.48\columnwidth]{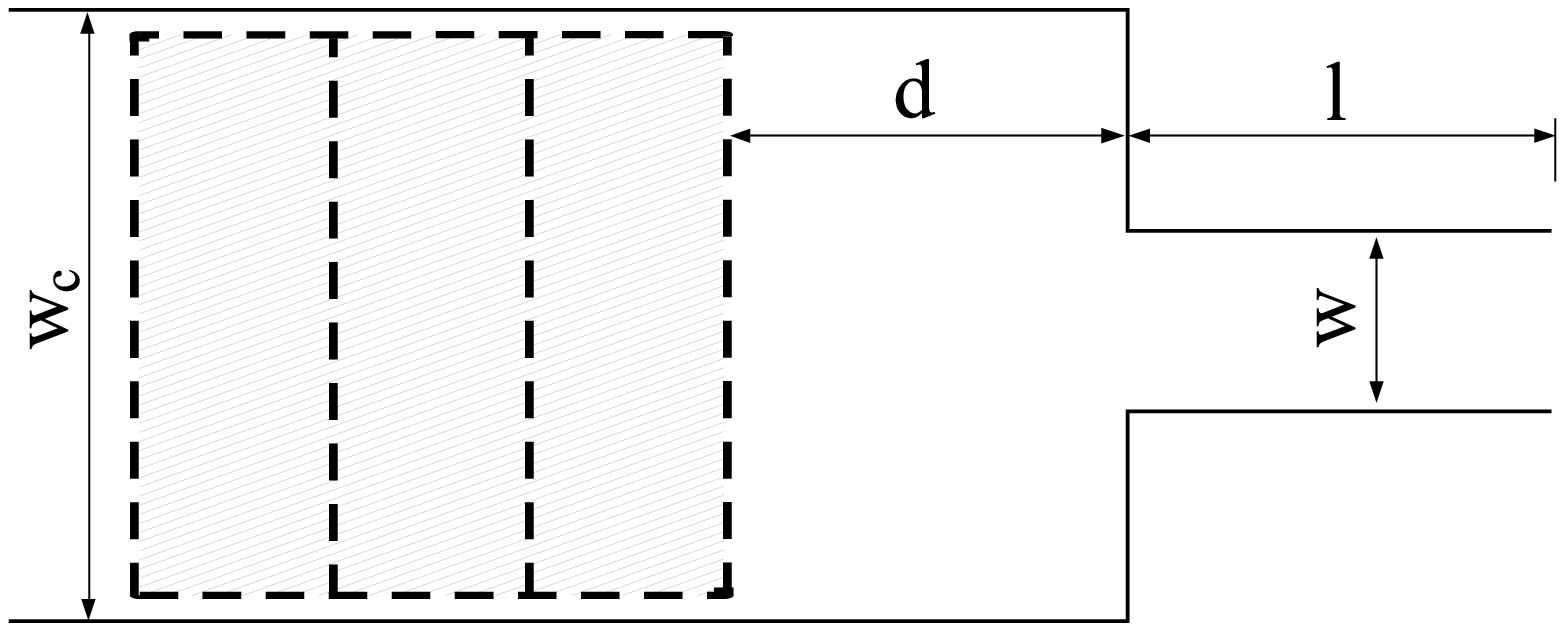}
  }
  \caption{Sketch of the experimental setup to determine the fundamental 
    diagram (left) and to analyze the flow through bottlenecks (rigth).
  }
  \label{EX-SETUP-SKETCH}
\end{figure}

The experiments were arranged 2006 in the wardroom of the `Bergische Kaserne
D\"usseldorf'. The group of test persons was composed of soldiers. 
Fig.~\ref{EX-SETUP-SKETCH} (left) shows a sketch of the experimental setup to 
determine the fundamental diagram. We performed runs for different widths $w$ 
as well as uni- and bidirectional flows. To scan the whole density regime the 
number of the pedestrians inside the corridor was changed. The right figure 
shows the sketch of the experimental setup to analyze the flow through 
bottlenecks. We performed runs for different bottleneck widths $w$, corridor 
widths $w_c$, bottleneck length $l$, number of pedestrians $N$ and distances 
to the entrance $d$. To ensure an equal initial density for every run, holding  
areas were marked on the floor (dashed regions). All together 99 runs with up 
to 250 people distributed over five days were performed. 

\section{Influence of the Measurement Method}
\label{meas-meth}

The discussion outlined in Sec.~\ref{sec-emp} is put into perspective
by two observations. First we note that in the majority of cases the
data come without fluctuations and error margins and thus, strictly
speaking, there is no contradiction. Second it is well known in
vehicular traffic that different measurement methods can lead to
deviations for the fundamental diagram
\cite{Leutzbach1972,Kerner2004}. In previous experimental studies of
pedestrian traffic, different kinds of measurement methods are used,
and often a mixture of time and space averages are realized due to
cost reasons. But in case of spatial and temporal inhomogeneities it
cannot be excluded that the averaging over different degrees of
freedom leads to non comparable results.
In this section we analyze how large the deviations due to different
measurement methods are. For this purpose we choose the most ordered and
controlled system examined in the project, namely the fundamental
diagram for the movement of pedestrians along a line under closed
boundary conditions during a stationary state. Due to the controlled
character of the movement it can be expected that deviations caused by
inhomogeneities give a lower bound for deviations in more disordered
systems.
 
\begin{figure}[thb]
  \centerline{
    \includegraphics[width=0.5\textwidth]{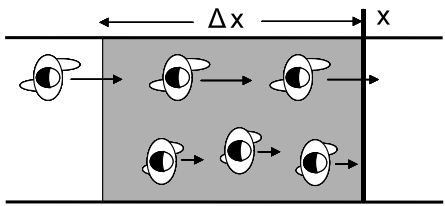}
  }
  \caption{Illustration of different measurement methods to determine the
    fundamental diagram. It has to be distinguished between local
    measurements at cross-section with position $x$ averaged over a
    time interval $\Delta t$ and measurements at certain time averaged
    over space $\Delta x$.}
  \label{SKIZZE-MES}
\end{figure}

In the following we introduce the basic quantities and the flow equation along
the measurement methods. The discussion follows the explanation in
text books for vehicular traffic \cite{Leutzbach1972,Kerner2004} and is adapted
to pedestrian characteristics. The sketch in Fig.~\ref{SKIZZE-MES} illustrates 
two principle possibilities to measure the observable like flow, velocity and 
density.  

{\bf Method A:} local measurement of the observable $O$ at a certain 
location $x$ averaging over a time interval $\Delta t$. We refer to 
this by $\langle O \rangle_{\Delta t}$. Measurements at a certain location
allow a direct determination of the flow $J$ and the velocity $v$.

\begin{equation}
  \langle J  \rangle_{\Delta t} = \frac{N}{\Delta t} = \frac{1}{\langle\Delta
    t_i \rangle_{\Delta t}} 
  \qquad \mbox{and} \qquad \langle v \rangle_{\Delta t}
  = \frac{1}{N} \sum_{i=1}^{N} v_{i} \,. 
  \label{FLOW-VEL}
\end{equation}
The flow is given as the number of persons $N$ passing a specified 
cross-section at $x$ per unit time. Usually it is taken as a scalar quantity
since only the flow normal to the cross-section is considered.  
To relate the flow with a velocity one measures the individual
velocities $v_{i}$ at location $x$ and calculates the mean value of
the velocity $\langle v \rangle_{\Delta t}$ of the $N$ pedestrians. In
earlier studies normally the velocity of a single pedestrian was
considered and only the number of pedestrians $N$ passing the
cross-section in the time interval $\Delta t$ are counted
\cite{Hankin1958,Togawa1955,Predtechenskii1978}. In principle it is
possible to determine the velocities $v_i$ and crossing times $t_i$ of
each pedestrian and to calculate the time gaps $\Delta
t_i=t_{i+1}-t_{i}$ defining the flow as the inverse of the mean value
of time gaps over the time interval $\Delta t$.

{\bf Method B} is to average the observable $O$ over space $\Delta x$ at
a specific time $t_k$ which gives $\langle O \rangle_{\Delta x}$.
The introduction of an observation area with extend $w \, \Delta x$
allows to determine directly the density $\rho$ and the velocity $v$:
\begin{equation}
  \langle \rho \rangle_{\Delta x} = \frac{N'}{w \, \Delta x} 
  \qquad \mbox{and} \qquad 
  \langle v \rangle_{\Delta x} = \frac{1}{N'} \sum_{i=1}^{N'} v_{i}\,. 
  \label{DENS-VEL}
\end{equation}

This method was used in combination with time-lapse photos. Often and 
due to cost reasons only the velocity of single pedestrians and 
the mean value of the velocity during the entrance and exit times were
considered \cite{Older1968,Navin1969}.

{\bf Flow equation:} To connect these methods and to change between 
different representations of the fundamental diagram the hydrodynamic 
flow equation $J=\rho\,v\,w$ is used. It is possible to derive the flow 
equation from the definition of the observables introduced above by using 
the distance $\Delta \tilde x = \Delta t \, \langle v \rangle_{\Delta t}$.
Thus one obtains 
\begin{equation}
  J = \frac{N}{\Delta t} = \frac{N}{\Delta \tilde x\;w} \frac{\Delta \tilde x
    \;w}{\Delta t} = \tilde \rho \; \langle v \rangle_x \; w \qquad \mbox{with}
  \qquad \tilde \rho = \frac{N}{\Delta \tilde x \; w}.
  \label{FLOW-EQUATION}
\end{equation}

At this point it is crucial to note that the mean values $\langle v
\rangle_x$ and $\langle v \rangle_t$ do not necessarily correspond.
This can already be seen by examination of Fig.~\ref{SKIZZE-MES}.
Thus a density calculated by $\tilde \rho = \langle J \rangle_{\Delta
  t}/\langle v \rangle_{\Delta t}$ may differ from a direct
measurement of the density via $\langle \rho \rangle_{\Delta x}$. We
come back to this point later.

\begin{figure}[thb]
  \centerline{
    \includegraphics[angle=-90, width=0.325\columnwidth]{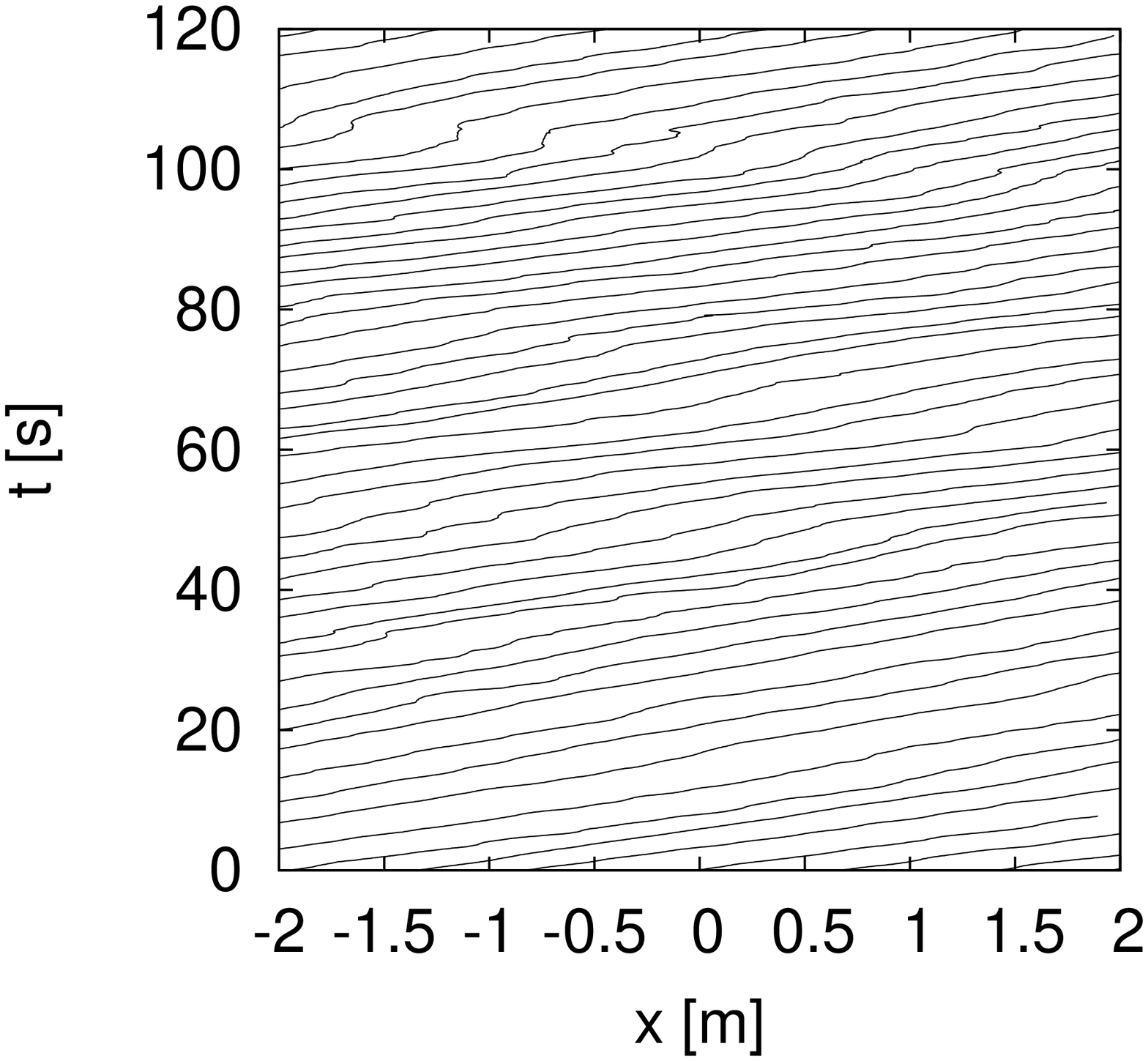}
    \includegraphics[angle=-90, width=0.325\columnwidth]{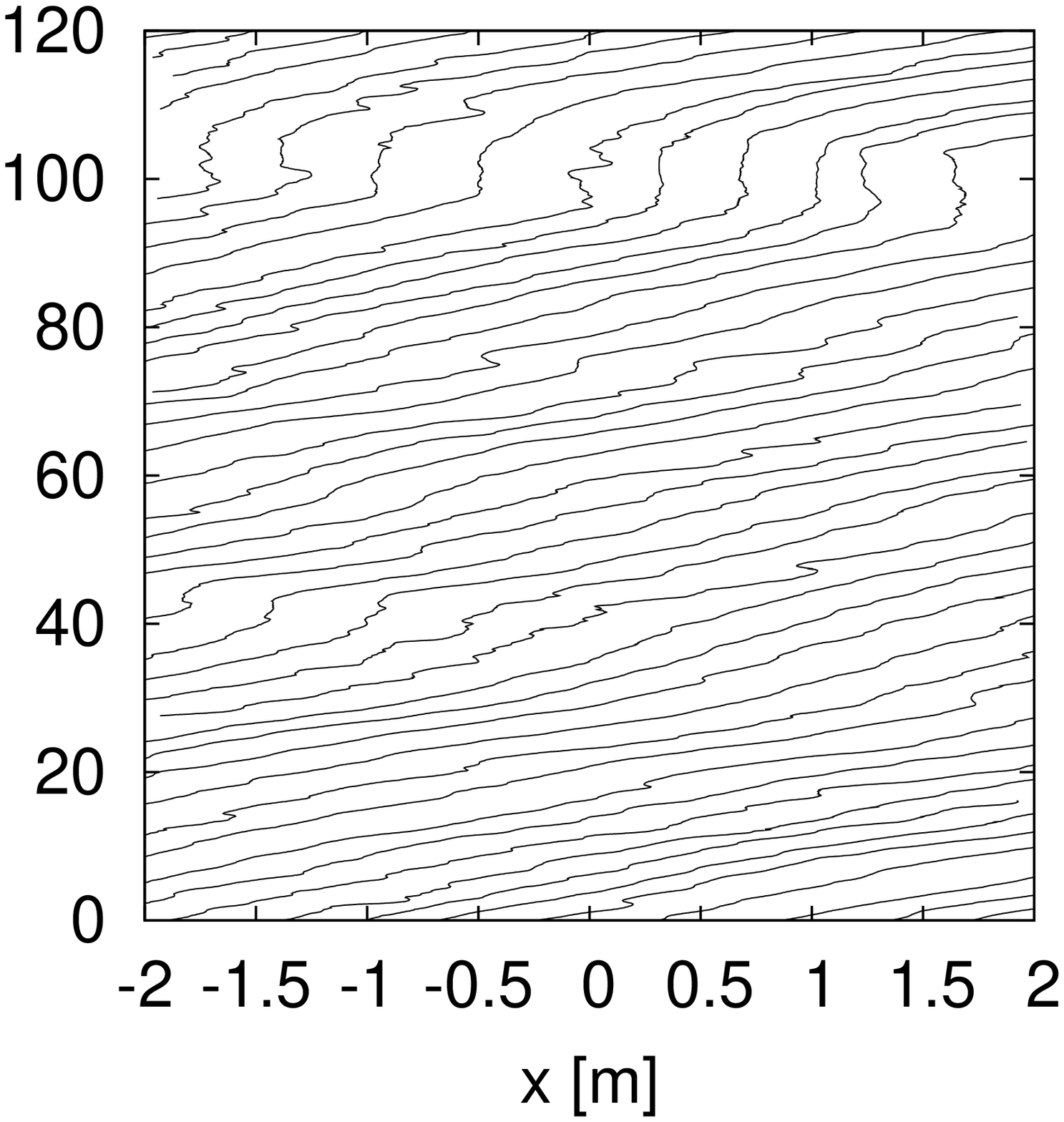}
    \includegraphics[angle=-90, width=0.325\columnwidth]{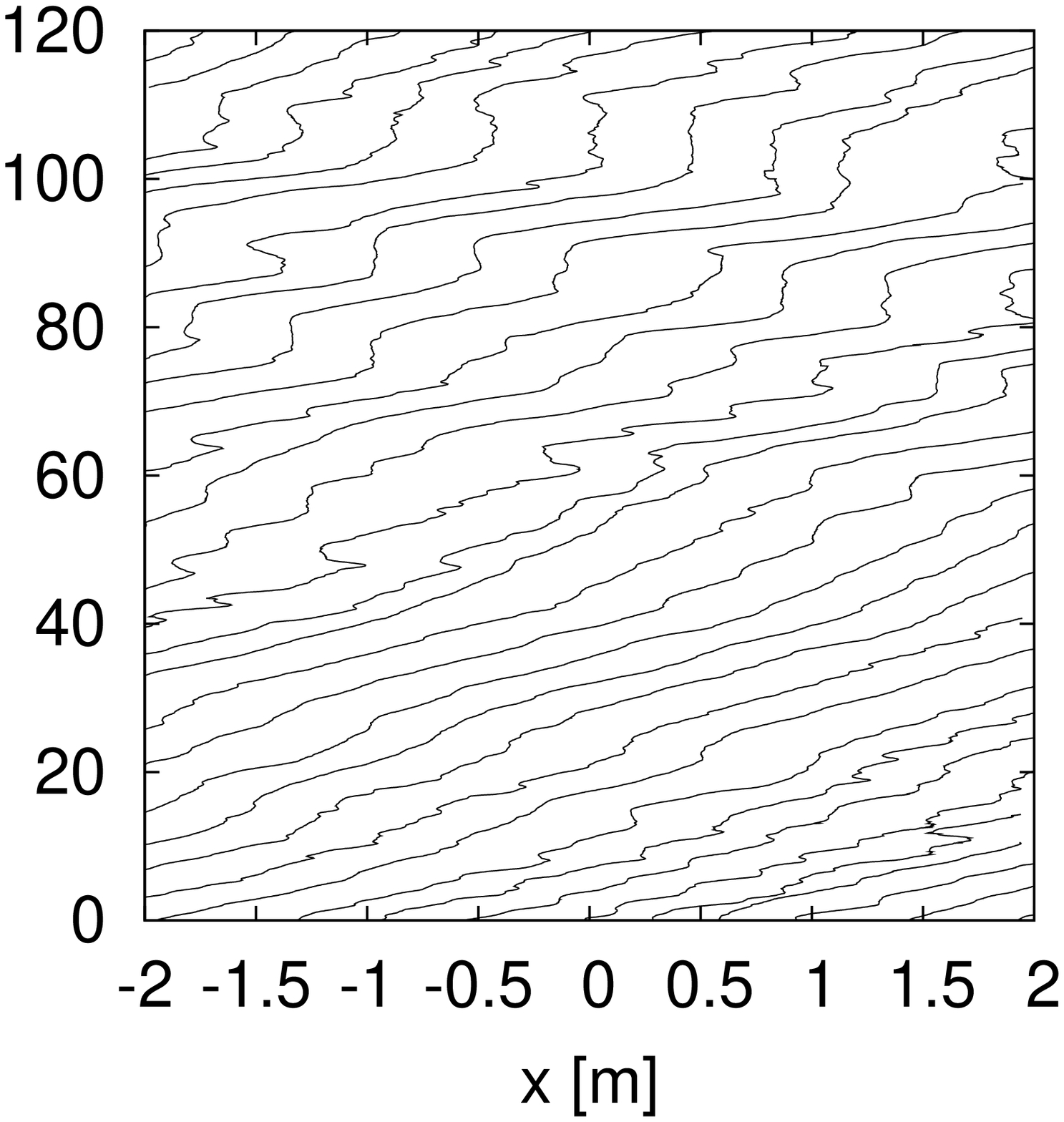}
  }
  \caption{Projection of the trajectories to the $(x,t)$-plane of the movement
    along a line for the runs with $N=45,56,62$ (from left to right). For
    increasing $N$ the dynamics becomes more unordered and the trajectories
    show intermittent stopping by a constant $x$-values in time.} 
  \label{TRAJ}
\end{figure}

As already mentioned above we choose the most simple system to get an
estimation for the lower bound of deviation resulting from different
measurement methods.  To measure the fundamental diagram of the
movement along a line we performed $12$ runs with varying number of
pedestrians, $N=17$ to $N=70$. Fig.~\ref{TRAJ} shows the projection
of the trajectories to the $(x,t)$-plane for the runs with $N=45, 56$
and $62$. For the movement along a line we set $w=1$ in the equations
introduced above. We note again that the different measurements shown
in the next figures are based on the same set of trajectories
determined automatically from video recordings of the measurement area
with high accuracy ($x_{\rm err} \pm 0.02m$). The data analysis is
restricted to the stationary state.

\begin{figure}[thb]
  \centerline{
    \includegraphics[angle=-90, width=0.42\columnwidth]{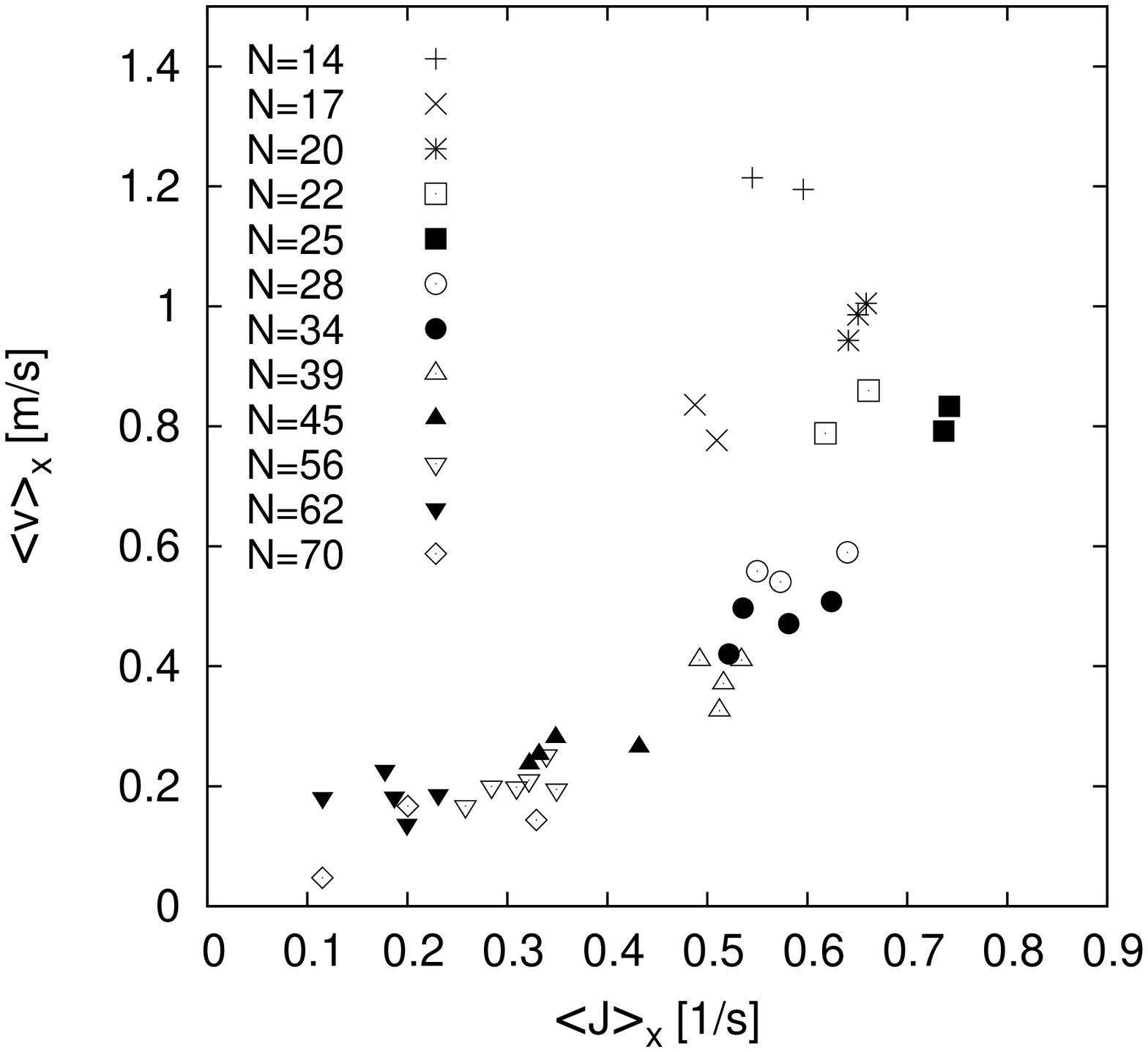}
    \hspace{0.5cm}
    \includegraphics[angle=-90, width=0.42\columnwidth]{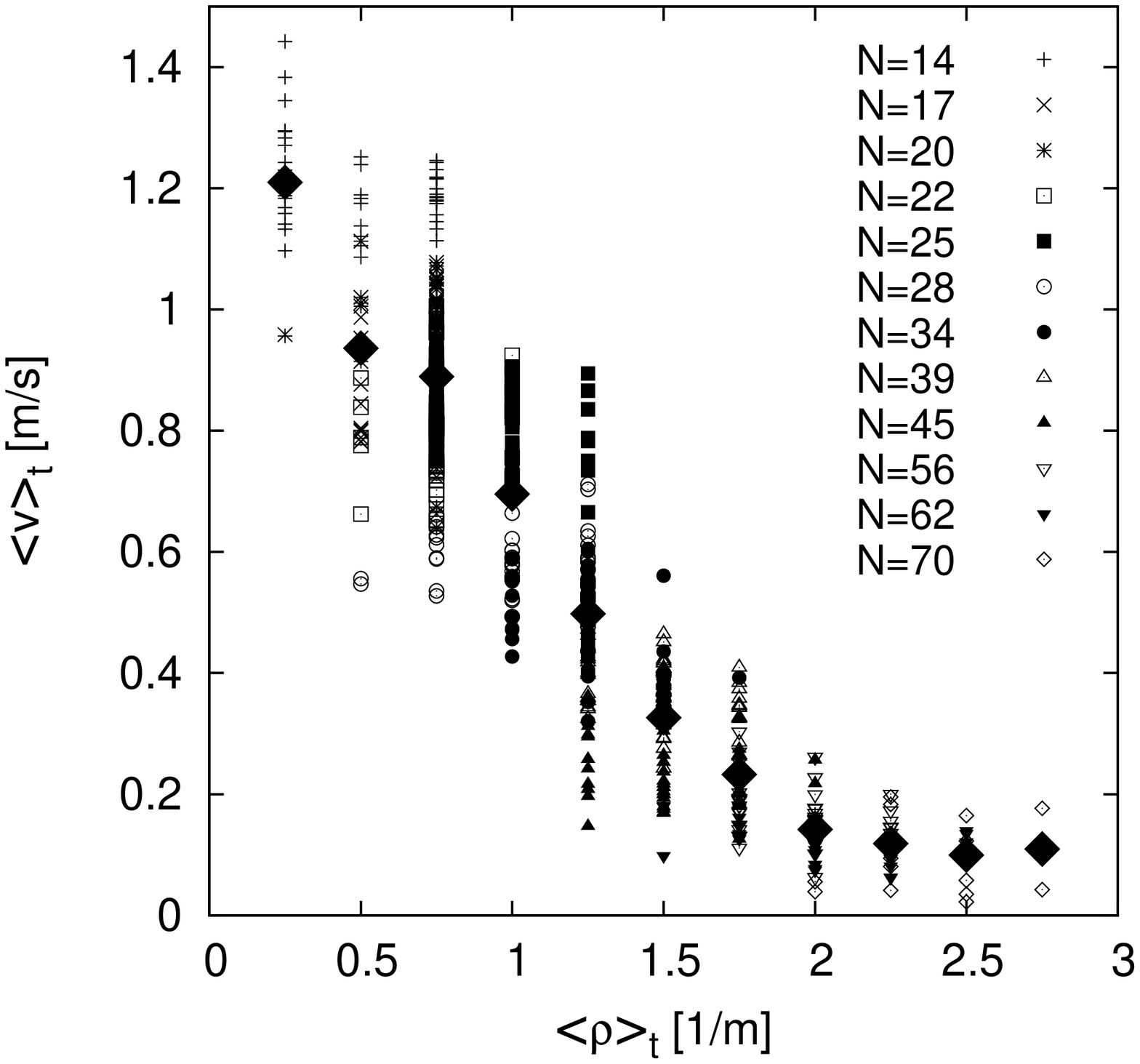}
  }
  \caption{Fundamental diagrams measured at the same set of trajectories but
    with different methods. Left: Measurement at a certain
    cross-section averaging over time interval (Method A). Right:
    Measurement at a certain point in time averaging over space
    (Method B). Large diamonds give the over all mean value of the
    velocity for one density value.}
  \label{FD-MESS-METH}
\end{figure}

Fig.~\ref{FD-MESS-METH} shows the direct measurements according to
Method A and B.  For Method A we choose the position of the
cross-section $x=0$ and a time interval of $\Delta t=30s$, see
Fig.~\ref{TRAJ}. For Method B the area ranges from $x=-2m$ to $x=2m$,
and we performed the averaging over space each time $t_k$ a pedestrian
crossed $x=0$. For Method B we note that the fixed length of the
observation area of $4m$ results in discrete density values with
distance $\Delta \rho = (4m)^{-1}$. For each density value large
fluctuations of the velocities $\langle v \rangle_t$ are observed. The
large diamonds in the right figure of \ref{FD-MESS-METH} represent the
mean values over all velocities $\langle v \rangle_t$ for one density.
The flow equation (\ref{FLOW-EQUATION}) allows to switch the direct
measurement of Method A and B into the most common representation of
the fundamental diagram $J(\rho)$.

\begin{figure}[thb]
  \centerline{
    \includegraphics[angle=-90, width=0.45\columnwidth]{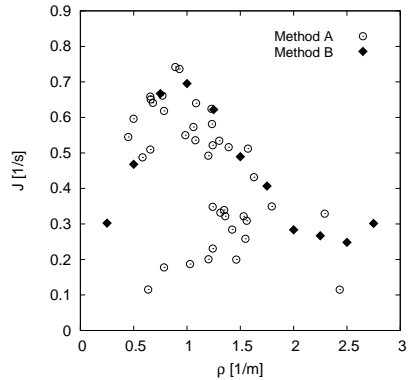}
  }
  \caption{Fundamental diagram determined by different measurement
    methods. {\em Method A}: Direct measurement of the flow and
    velocity at a cross-section. The density is calculated via $\rho
    =\langle J \rangle_{\Delta t}/ \langle v \rangle_{\Delta t}$. {\em
      Method B}: Measurement of the density and velocity at a certain
    time point averaged over space. The flow is given by $J=\rho \,
    \langle v \rangle_{\Delta x}$.  }
  \label{RESULT}
\end{figure}

Fig.~\ref{RESULT} shows a comparison of fundamental diagrams using the
same set of trajectories but different measurement methods. In
particular for high densities, where jam waves are present, the
deviations are obvious. This is in agreement with
Fig.~\ref{RHO-JS-EMP} where almost all curves agree for low densities
and disagree for high densities. For the high density regime the
trajectories show inhomogeneities in time and space, which do not
correspond, see Fig.~\ref{TRAJ}. The averaging over different degrees
of freedom, the time $\Delta t$ for Method A and the space $\Delta x$
for Method B lead to different distribution of individual velocities.
Thus one reason for the deviations is that the mean values of the
velocity measured at a certain location by averaging over time do not
necessarily conform to mean values measured at a certain time averaged
over space. However, the straightforward use of the flow equation
neglects these differences.  In \cite{Leutzbach1972} it was stated
that the difference can be cancelled out by using the harmonic average
for the calculation of the mean velocity for Method A. We test this
approach and found that the differences do not cancel out and the data
are only in conformance if one takes into account the fluctuations and
calculates the mean velocity by the harmonic average. But for states
where congestions lead to an intermittent stopping, fluctuations of
the density measured with Method A are extremly large and can span
over the whole density range observed. This belongs to the fact that
in Method A the density is determined indirectly by calculating
$\tilde \rho =\langle J \rangle_{\Delta t}/ \langle v \rangle_{\Delta
  t}$. In the high density range the flow as well as the velocity have
small values causing high fluctuations for the calculated density.

\section{Conclusions}

This contribution summarizes open questions and differences concerning
specifications of the fundamental diagram and bottleneck flow in the
literature.  In particular for the high density regime of the
flow-density relation the discrepancies are not negligible. For the
flow through bottlenecks it is an open question, why the maximal flow
values through bottleneck exceed significantly the maxima of the
fundamental diagrams. To dissolve these discrepancies we performed
laboratory experiments with up to 250 people. The trajectories of each
pedestrian are determined with high accuracy. As a first step of the
analysis we investigated how the way of measurement influence the
resulting relations. Surprisingly we found that even for the most
regular and simplest system, namely the movement of pedestrians along
a line under periodic boundary conditions, large deviations result if
different measurement methods are applied. The reason for this is the
averaging over different degrees of freedom in a discrete system with
large inhomogeneities. Thus it cannot be excluded that the deviations
discussed in Sec.~\ref{sec-emp} result from different measurement
methods amongst other causes. This statement is supported by the
observation that the Fig.~\ref{RHO-JS-EMP} where allmost all curves
agree for low densities and disagree for high densities. For a
systematic study and a meaningful discussion of the influence of
culture or the changing population demographics on pedestrian
characteristic it is necessary to assure that the studies compared are
based on the same measurement approach. This applies accordingly for
the validation of model results with experimental data.


\section*{Acknowledgements}
We would like to thank the German Science Foundation (DFG) for funding
this project under DFG-Grant No. KL 1873/1-1 and SE 1789/1-1.
A.~Seyfried and A.~Schadschneider are grateful to the ped-net-group
({\tt www.ped-net.org}) for intensive and inspiring discussions.



\end{document}